\documentclass[twocolumn]{aastex63}
\usepackage{amsmath}
\usepackage{hyperref}
\usepackage{graphicx}
\usepackage{booktabs}
\usepackage{diagbox}
\usepackage{float}    

\usepackage{xcolor}
\usepackage{bold-extra}

\shorttitle{LRD Properties \& Evolution}
\shortauthors{Pang et al.}

\begin{document}

\title{The Structure and Evolution of LRDs: Insights from JWST NIRSpec Medium and High Resolution Spectroscopy at $z\sim4$}

\author[0009-0005-3823-9302]{Yuxuan Pang}
\affiliation{School of Astronomy and Space Science, University of Chinese Academy of Sciences (UCAS), Beijing 100049, China}

\author[0000-0002-9373-3865]{Xin Wang}
\affiliation{School of Astronomy and Space Science, University of Chinese Academy of Sciences (UCAS), Beijing 100049, China}
\affiliation{National Astronomical Observatories, Chinese Academy of Sciences, Beijing 100101, China}
\affiliation{Institute for Frontiers in Astronomy and Astrophysics, Beijing Normal University, Beijing 102206, China}

\author[0000-0003-0202-0534]{Cheng Cheng}
\affiliation{Chinese Academy of Sciences South America Center for Astronomy, National Astronomical Observatories, CAS, Beijing 100101, People’s Republic of China}
\affiliation{Key Laboratory of Optical Astronomy, NAOC, 20A Datun Road, Chaoyang District, Beijing 100101, People’s Republic of China}

\author[0009-0007-6655-366X]{Shengzhe Wang}
\affiliation{National Astronomical Observatories, Chinese Academy of Sciences, Beijing 100101, China}
\affiliation{School of Astronomy and Space Science, University of Chinese Academy of Sciences (UCAS), Beijing 100049, China}

\author[0009-0004-7133-9375]{Hang Zhou}
\affil{School of Astronomy and Space Science, University of Chinese Academy of Sciences (UCAS), Beijing 100049, China}

\author[0009-0006-1255-9567]{Qianqiao Zhou}
\affiliation{School of Astronomy and Space Science, University of Chinese Academy of Sciences (UCAS), Beijing 100049, China}

\author[0000-0002-7350-6913]{Xue-Bing Wu}
\affiliation{Department of Astronomy, School of Physics, Peking University, Beijing 100871, People's Republic of China}
\affiliation{Kavli Institute for Astronomy and Astrophysics, Peking University, Beijing 100871, People's Republic of China}

\author[0000-0002-3254-9044]{Karl Glazebrook}
\affiliation{Centre for Astrophysics and Supercomputing, Swinburne University of Technology, Hawthorn, VIC 3122, Australia}

\correspondingauthor{Yuxuan Pang, Xin Wang, Cheng Cheng}
\email{pangyuxuan@ucas.ac.cn}
\email{xwang@ucas.ac.cn}
\email{chengcheng@nao.cas.cn}

\begin{abstract}
We present an analysis of medium/high-resolution JWST/NIRSpec spectra for 11 LRDs at $z \sim 4$. By decomposing the broad and narrow components of the Balmer emission lines, we investigate the connection between line emission and UV/optical continua for the LRD population. We find that the broad H$\alpha$ luminosity strongly correlates with the optical continuum (but not with the UV), indicating a common AGN origin for both. In contrast, the [O~\textsc{III}] line strength is correlated with the UV continuum rather than the optical. Using the width and luminosity of the broad H$\alpha$ line, we estimate central black hole masses of $10^6$–$10^8\ M_{\odot}$ accreting at high Eddington ratios ($\lambda_{\mathrm{Edd}} \sim 0.6$), consistent with an early, rapid-growth phase of AGN evolution. Assuming a constant mass accretion rate in the framework of slim-disk models, we infer growth timescales of $\sim10^5-10^7\rm yr$, and suggest LRDs may evolve into narrow-line Seyfert 1 galaxies. Upper limits from our spectra indicate that LRDs exhibit intrinsically weak optical Fe~\textsc{II} emission compared to typical AGN. To simultaneously account for the inferred broad-line region size and observed luminosity, we propose a ``Clumpy Envelope'' model in which the optical emission arises from an extended, clumpy gas with a characteristic radius of tens of light-days. The diversity in observed optical continuum shapes can be explained by radial temperature gradients and self-absorption effects within this structure. Our results demonstrate the power of JWST high-resolution spectroscopy in probing the central engines and physical nature of the LRD population.
\end{abstract}


\section{Introduction} \label{sec: intro}
Little Red Dots (LRDs) are a class of luminous, high-redshift objects and represent one of the most significant discoveries of the early years of JWST operations. Their high number density, particularly in the high-redshift universe \citep[e.g.][]{2024ApJ...963..129M,2025ApJ...986..126K}, suggests potential effects on supermassive black hole and galaxy co-evolution. Furthermore, statistical analysis of LRDs may help to refine our understanding of the black hole seed formation process \citep{2025ApJ...989L...7T,2026ApJ...996L..19J}.

Observationally, LRDs exhibit distinctive characteristics in the optical to near-infrared bands. Their defining properties include: 1) a distinctive "V-shaped" continuum, characterized by a blue UV slope and an apparent change in slope near the Balmer break \citep{2025ApJ...995..118S,2025ApJ...991...37A}; 2) a compact morphology, unresolved by JWST at least by NIRCam long-wavelength channels, implying sizes of $\lesssim 100 ~\rm pc$ \citep{2024ApJ...968...38K,2026arXiv260106015Y}; 3) broad hydrogen and helium emission lines with widths of larger than several thousand $\rm km~s^{-1}$ \citep{2024ApJ...964...39G,2024ApJ...963..129M,2026MNRAS.tmp...94J}; and 4) a high incidence of red- and blue-shifted Balmer and He~\textsc{I} absorption features \citep{2024MNRAS.535..853J,2024ApJ...974..147L,2025ApJ...986..126K}.

The intrinsic physical origin of the LRD spectral energy distribution (SED) has been extensively debated due to the high degeneracy between active galactic nuclei (AGN) and stellar components. Initial interpretations proposed LRDs as either dust-reddened AGN \citep[e.g.,][]{2024ApJ...963..129M,2025ApJ...978...92L,2025ApJ...986..126K} motivated by the discovery of broad Balmer lines in a high fraction of LRDs \citep[e.g.,][]{2024ApJ...964...39G,2025A&A...702A..57H}, or massive post-starburst or dusty star-forming galaxies \citep[e.g.,][]{2023Natur.616..266L,2024ApJ...968....4P,2024ApJ...969L..13W,2024ApJ...977L..13B}, due to the frequent detection of Balmer breaks \citep{2025ApJ...995..118S}. However, with expanding LRD samples and multi-wavelength follow-up observations, both models face significant challenges. The dusty AGN model fails to explain the general lack of mid- to far-infrared emission from cold and hot dust \citep[e.g.,][]{2024ApJ...975L...4C,2025ApJ...990L..61C,2025ApJ...991...37A,2025ApJ...991L..10S,2025A&A...700A.231X}, as well as the weakness of X-ray and radio emission expected from typical AGN \citep[e.g.,][]{2024ApJ...974L..26Y,2024ApJ...969L..18A,2025MNRAS.538.1921M,2024arXiv241204224M,2025ApJ...986..130G}. Conversely, purely stellar models struggle to reproduce the SEDs of the most extreme LRDs discovered to date \citep{2025A&A...701A.168D,2025arXiv250316596N}.

A promising alternative model interprets LRDs as rapidly accreting SMBHs embedded within dense, ionized gaseous envelopes \citep{2025arXiv250316595R,2025ApJ...980L..27I,2025arXiv250316596N}. This ``gas shell" framework can simultaneously account for several puzzling features, including the Balmer absorption lines, unusually large Balmer decrements, deep Balmer breaks, the large equivalent widths (EW) of H$\alpha$ emission, and the possible exponential profiles of broad-line components \citep[e.g.,][]{2024MNRAS.535..853J,2024ApJ...974..147L,2025MNRAS.544.3900J,2025A&A...701A.168D}. While the existence of the dense gas component is now widely accepted, the functions of that component are different among models. The model of \cite{2025ApJ...994..113L} explained the observed continuum SEDs by super-Eddington accreting black holes within massive envelopes. Theoretical work leveraging the ``quasistar" concept posits that LRDs may represent a stable evolutionary phase lasting tens of millions of years when radiation pressure could support the gas around the black hole \citep{2026ApJ...996...48B,Kohei2025model}, with their red optical spectra arising from an optically thick photosphere with temperatures of $5000–7000~\rm K$ and radius of $\sim 10^3~\rm AU$, analogous to stars near the Hayashi limit \citep{2025MNRAS.544.3407K}. Recently, updated non-spherical ``cocoon" models have been proposed to explain both the continuum and emission-line shapes through detailed radiative transfer \citep{2026arXiv260118864S}.

Although the models above could explain most of the observation features, key open questions remain on the structure of the gas (e.g., the location of broad-line regions (BLRs), the total size of the structure) and the properties of the central engines (e.g., normal AGN accretion or quasistars). To address these questions, recent studies have begun mapping the parameter space, including emission-line and continuum properties \citep[e.g.,][]{Anna2025c,2025arXiv251215853B,Asada26}.
\cite{Anna2025c} suggest a correlation between Balmer decrement, total luminosity of H$\alpha$ emission line, and continuum luminosity of $5100~\rm \AA$. High [O~\textsc{III}] EWs and the relations between [O~\textsc{III}] and UV magnitude are also reported. Moreover, the modified blackbody fitting shows the diversity of the LRD optical continuum. \cite{2025arXiv251215853B} focuses more on the continuum correlations between UV and optical bands and proposed evolution progresses of LRDs on the schematic diagram. However, these analyses only used low-resolution NIRSpec/prism spectra, which limit the ability to decompose broad and narrow emission-line components accurately. Besides, while LRDs are increasingly viewed as a new class of AGN, their precise relationship to classical AGN remains an open and challenging topic.

In this paper, we advance the understanding of LRDs by analyzing JWST NIRSpec medium/high-resolution spectra. We decompose the broad and narrow components of Balmer lines to investigate the correlations between line luminosity and the UV/optical continuum. This allows us to place new observational constraints on LRD models. Furthermore, we examine the location and potential evolutionary trajectories of LRDs on traditional AGN diagnostic diagrams, using physical properties derived from the broad H$\alpha$ component. The paper is structured as follows: Section \ref{sec: data} describes the sample and data. Section \ref{sec: fitting} details the spectral fitting methodology. Section \ref{sec: results} presents the main results of our work. Section \ref{sec: discuss} discusses the systematic uncertainties in black hole mass estimation, examines the black hole mass to stellar mass relation, and proposes a thicker gas envelope model to explain the diversity of optical continuum shapes. Section \ref{sec: summary} summarizes our conclusions. Throughout this work, we assume a $\Lambda$CDM cosmology with parameters $\Omega_m$ = 0.30, $\Omega_{\Lambda}$  = 0.7 and $h_0$ = 70 km $\text{s}^{-1}$ $\text{Mpc}^{-1}$.

\section{Data and Sample} \label{sec: data}
\subsection{Data}
We utilize all publicly available NIRSpec MSA spectra from version 4.4 of the DAWN JWST Archive \citep[DJA v4.4;][]{2025zndo...15472353}. The spectra were reduced using msaexp \citep{2023zndo...8319596B}, largely following the methodology outlined in \cite{2025A&A...693A..60H} and \cite{2025A&A...697A.189D}, with local background subtraction performed on nodded exposures where possible.

To further decompose the broad and narrow emission-line components, we carry out a detailed analysis using medium- and high-resolution NIRSpec spectra. These include data obtained with the G140M, G235M/H, and G395M/H gratings/filters from the RUBIES \citep{2025A&A...701A.168D}, JADES \citep{2023arXiv230602465E, 2025arXiv251001033C, 2025arXiv251001034S}, and NIRSpec GTO-Wide \citep{2024A&A...689A..73M}.

\subsection{Sample}
We adopt the LRD catalog from \cite{Anna2025c} for further analysis of related properties, which contains 116 LRDs. Briefly, \cite{Anna2025c} conducted a comprehensive selection of LRD sources using prism spectra from multiple programs and performed careful analysis of spectral line and continuum properties through fitting of UV/optical continua and emission lines.
To ensure that each source has a well-defined optical continuum shape and to obtain a more robust constraint on its optical luminosity, we require that the peak flux density in the rest-frame optical band is clearly detected in the prism spectra from the \cite{Anna2025c} catalog, this requires the maximum wavelength of the prism spectrum ($5.3\mu m$) is larger than the 3$\sigma$ upper limit of the peak flux defined in \cite{Anna2025c}, which is related to a $z\lesssim4$ for a blackbody with $5000~\rm K$. Such a step leaves 38 sources. We then cross-match the remaining sample with all DJA datasets that contain medium or high-resolution observations of the H$\alpha$ emission line and yield 13 individual sources. To further separate the broad and narrow emission-line components, we further confirmed that all the 13 sources have H$\alpha$ peak signal-to-noise ratio (SNR) larger than 3. Among 13 sources, 2 show no significant broad component (full width at half maximum, $\rm FWHM > 600 km/s$) in their H$\alpha$ emission lines (See details of the fitting method in the Section \ref{sec: fitting}). Of the remaining 11 sources, four also have medium-resolution observations covering the spectral region around H$\beta$ and [O~\textsc{III}], three of which have a broad component in the H$\beta$ emission line. Compared to the parent sample of 116 LRDs from \cite{Anna2025c}, our final sample of 11 sources has a lower redshift but similar distributions in optical luminosity, UV magnitude, and H$\alpha$ flux.

For the continuum properties, we adopt the results from \cite{Anna2025c}\footnote{\url{10.5281/zenodo.17665942}}. Briefly, \cite{Anna2025c} gives the continuum fitting of the prism spectra in two parts separated by the Balmer limit. The rest-frame UV continuum was fitted with a power law, and the fitting profile was extrapolated to a rest-frame wavelength of $1500~\rm \AA$ to compute the UV magnitude ($M_{\rm UV}$). For the optical band, a modified blackbody spectrum was used for fitting, and its integration yielded the optical luminosity ($L_{\rm opt}$).

\section{Spectra Fitting Process} \label{sec: fitting}
After downloading the NIRSpec 1D spectra from the DJA, we began the emission line fitting process with continuum subtraction. The continuum was fitted in nearby wavelength windows on both sides of each emission line: for H$\alpha$ we used $6264–6414\rm \AA$ and $6714–6864\rm \AA$; for H$\beta$ and [O~\textsc{III}] we used $4561–4711\rm \AA$ and $5157–5307\rm \AA$. Within these windows, the continuum was modeled with a quadratic polynomial. A quadratic function was chosen because the rest-frame optical continua of LRDs exhibit a convex blackbody-like shape, which is better approximated locally by a second-order polynomial; however, switching to a linear or power-law function does not significantly affect the final line-fitting results.

After subtracting the continuum, we used the \texttt{LMFIT} package\footnote{\url{10.5281/zenodo.16175987}} to perform multi-component Gaussian fitting of H$\alpha$, H$\beta$, and [O~\textsc{III}] within specified wavelength ranges: $6414–6714 \rm \AA$ for H$\alpha$ and $4711–5157 \rm \AA$ for H$\beta$ and [O~\textsc{III}]. Following \cite{2025arXiv250520393Z}, our selection criterion for the broad component is $\rm{FWHM}\geq600~km/s$. This criterion is more lenient than the often adopted criterion of $\geq1000~\rm km/s$ \citep[e.g.,][]{2024ApJ...964...39G,2024ApJ...963..129M}, and similar with several works adopting an even smaller FWHM threshold \citep[e.g.,][]{2024A&A...691A.145M, 2025ApJ...986..165T, 2026MNRAS.tmp...94J}. For H$\alpha$, which has relatively high SNR, we adopted a model consisting of two broad components and one narrow component. Multiple Gaussian components to fit the H$\alpha$ emission line profile shows enough flexibility of broad H$\alpha$ line in AGN studies \citep[e.g.,][]{2022ApJS..263...42W,2023ApJS..265...25J}. Recent work has shown that multiple Gaussians can also adequately recover long-tailed line wings \citep{scholtz2026littleredbluedots}. We adopted two broad Gaussian components to strike a balance between the resolution of the JWST M/H spectra and a reasonable number of free parameters. We find that 11 of 13 sources have H$\alpha$ broad component(s) detection, with a total flux $\rm SNR>3$ (the remaining two sources are srcid-2794 and srcid-4273, with $\rm SNR=2.1$ and $<0.1$, respectively). For H$\beta$, whose broad component is often fainter, we used one broad and one narrow Gaussian and suggest that three out of four spectra have an H$\beta$ broad component with $\rm SNR>3$. It's worth noticing that the velocity offsets for all H$\alpha$ and H$\beta$ components are less than $\rm 250~km~s^{-1}$, and the median value of the broad H$\alpha$ and H$\beta$ components is $\rm 106~km~s^{-1}$, which is much smaller than the FWHM and broadly consistent with typical AGN broad line offsets \citep[e.g., $\rm \sim 220~km~s^{-1}$ from][]{2022ApJS..263...42W}.

For [O~\textsc{III}], we fixed the relative velocities of the $\lambda4959$ and $\lambda5007$ lines and fitted them with a single narrow component each. The fitted [O~\textsc{III}] line ratio falls between 2.8 and 3.2 with a typical error of 0.15, consistent with the theoretical value of 2.98, indicating that fixing the ratio does not significantly influence the fitting results. Therefore, we accepted the [O~\textsc{III}] flux without further imposing a fixed flux ratio fitting. 
To investigate potential [O~\textsc{III}] outflow in our sample, we performed a comparative fitting by adding an additional set of [O~\textsc{III}] doublet Gaussian components with the wavelength ratio fixing, and required the outflow component to have $\rm FWHM>300~km~s^{-1}$ and a peak SNR greater than 3 \citep{2025ApJ...984..182X}. By comparing BIC between fits with and without the outflow component, we found that no source yields $\rm \Delta BIC <-6$, indicating no significant detection of [O III] outflows.

Some of the sources in our sample exhibit absorption features in the H$\alpha$ and H$\beta$ emission lines. To systematically identify these features, we added an additional Gaussian absorption component to each H$\alpha$ and H$\beta$ model and evaluated the resulting improvement in the BIC. Adopting a threshold of $\rm \Delta BIC>10$, we find that srcid-28074, srcid-28812, and srcid-154183 show absorption in H$\alpha$, while srcid-28074 also shows absorption in H$\beta$. 
Adding absorption components enlarges the parameter space, which may cause the fitting to become trapped in local minima when spectra are fit after Monte Carlo (MC) sampling. Therefore, to investigate the fitting errors, we used an iterative approach rather than simultaneously fitting the emission and absorption features in these sources. We first masked the absorption regions and fitted the emission lines with multiple Gaussian components. Then we compared the fitted emission profile to the observed data, and fitted the absorption component with a single Gaussian profile. This process was iterated three times; during each iteration, the absorption profile obtained from the previous step was added to the observed data when fitting the emission lines. 
The fitting has converged for our sample since the difference in $\chi^2$ between the second and third iterations was less than 0.01 for all spectra.

\begin{figure*}
\includegraphics[width=1.0\textwidth]{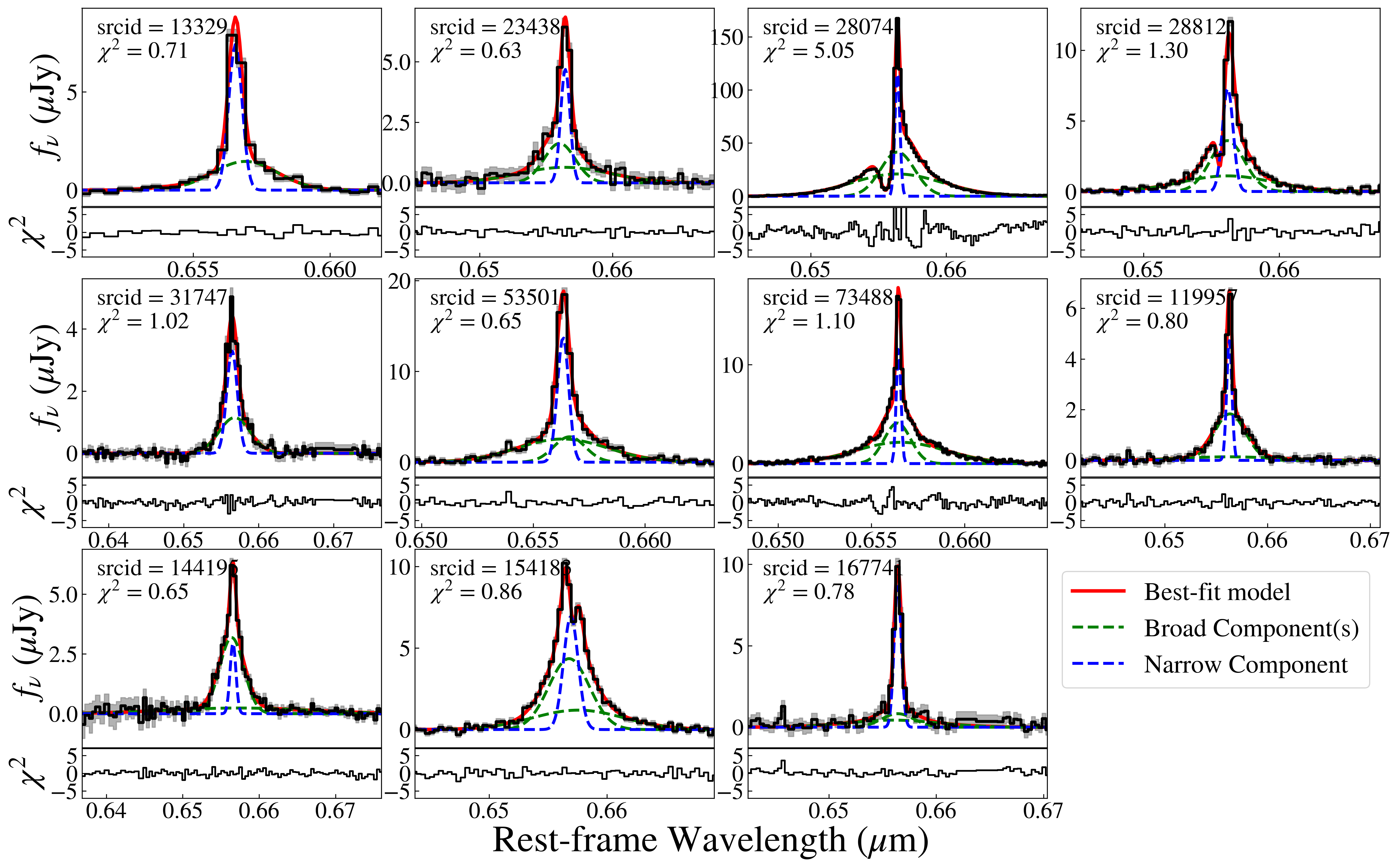}
\caption{Spectral fitting results for the H$\alpha$ emission lines in our LRD sample. In each panel, the black histogram represents the observed 1D flux density $f_{\nu}$ (in units of $\mu \rm Jy$), the gray shaded region indicates its 1$\sigma$ uncertainty, and the best fitting total model is shown as the solid red line. The broad and narrow emission-line components are depicted by the green and blue dashed curves, respectively. The source ID and reduced $\chi^2$ results are indicated in each panel. The lower panels show the residuals of each fitting.
\label{fig: ha_fitting}}
\end{figure*}

\begin{figure*}
\includegraphics[width=1.0\textwidth]{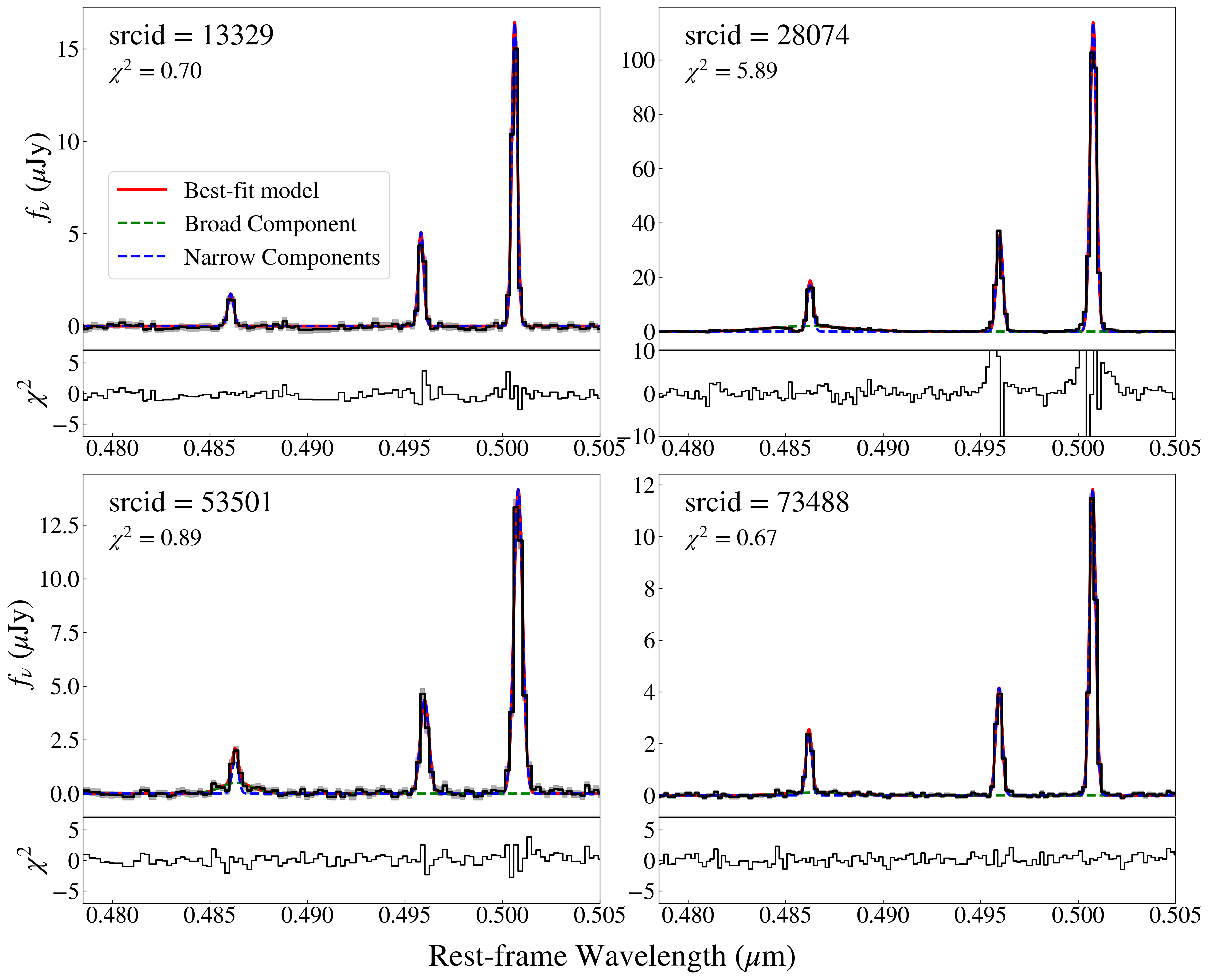}
\caption{Spectral fitting results for the H$\beta$ and [O~\textsc{III}] emission lines in our LRD sample. The line components, labels, and annotations are the same as Figure \ref{fig: ha_fitting}. 
\label{fig: hb_fitting}}
\end{figure*}

The resulting fits (including the absorption features) for H$\alpha$ and for H$\beta$/[O~\textsc{III}] are shown in Figure \ref{fig: ha_fitting} and Figure \ref{fig: hb_fitting}, respectively, together with the individual broad, narrow Gaussian components. For all sources except srcid-28074, the reduced $\chi^{2}$ values of fitting fall within the 95\% confidence interval, except for the H$\alpha$ profile of srcid-23438, srcid-53501, and srcid-144195, which fall within the 95\%–99\% interval, suggesting a slight overfitting. We note that the reduced $\chi^{2}$ of srcid-28074 is higher than expected, likely because a single Gaussian absorption profile may not fully capture the complexity of the absorption features in this high SNR spectrum. Nevertheless, all the fitting results show high consistency in the line profile and minimal total flux difference between the best‑fit model and the observed data.

We compute the line properties based on the best-fitting model. For the broad H$\alpha$ component, we combine the two Gaussian components and measure the FWHM and total flux from the resulting profile. For the narrow H$\alpha$ component and the broad/narrow H$\beta$ components, we adopt the FWHM and flux directly from the corresponding single Gaussian fits. For the [O~\textsc{III}] doublet, we derive the line flux from the fitted Gaussian model. The instrumental broadening correction was applied using the wavelength-dependent resolution of NIRSpec, with revisions on the resolution of medium/high-resolution spectra ($R\sim1000$ for medium-resolution \& $R\sim3000$ for high-resolution) \citep{2025A&A...702L..12S}. For each spectrum, we also performed an MC simulation using the error array to estimate parameter uncertainties. All derived total H$\alpha$ luminosities ($L_{\rm H\alpha,~total}$) and [O~\textsc{III}] $\lambda5007$ luminosities ($L_{\rm [O III],~5007}$)
agree within 10\% with the results from prism spectra in \cite{Anna2025c}, as expected given differences in continuum model and spectral resolution between prism and medium/high-resolution data.

Since our FWHM is derived from the combination of all broad components and captures the line shape, an exponential profile may further reduce the intrinsic FWHM. We will discuss the comparison results in Section \ref{subsec: mbh_disuss}. Notably, none of the four medium-resolution spectra covering H$\beta$ and [O~\textsc{III}] show any detectable Fe~\textsc{II} features. We therefore derived 3$\sigma$ upper limits for the Fe~\textsc{II} flux by fitting an empirical Fe~\textsc{II} template by \cite{2004A&A...417..515V} to the 1D error spectrum. Following the Fe~\textsc{II} strength ($R_{\rm FeII} = F_{\rm Fe}/F_{\rm H\beta}$) definition used in \cite{2022ApJS..263...42W} and \cite{2025ApJ...987...48P}, we computed the upper limit for the Fe~\textsc{II} flux integrated over the rest-frame wavelength range $4434–4684\rm \AA$. The information on emission line fitting results is listed in Table \ref{tab: info_ha} and Table \ref{tab: info_hb}.

\begin{deluxetable*}{lcccccc}
\tablecaption{\centering The properties of LRDs in our sample
\label{tab: info_ha}}
\tabletypesize{\footnotesize}
\tablehead{
\colhead{srcid} & \colhead{Redshift} & \colhead{$\rm FWHM_{\rm H\alpha}$} & \colhead{$\log L_{\rm H\alpha,~broad}$} & \colhead{$\log M_{\rm BH}$} & \colhead{$\lambda_{\rm Edd}$} & \colhead{$\log M_{*}$} \\
 &  & \colhead{$\rm km~s^{-1}$} & $\log(\rm erg~s^{-1})$ & \colhead{$\log(M_{\odot})$} &  & \colhead{$\log(M_{\odot})$} \\
\colhead{(1)} & \colhead{(2)} & \colhead{(3)} & \colhead{(4)} & \colhead{(5)} & \colhead{(6)} & \colhead{(7)}
}
\startdata
13329 & 3.94 & 814 $\pm$ 33 & 42.29 $\pm$ 0.03 & 6.41 $\pm$ 0.04 & 0.36 $\pm$ 0.04 & 9.32 $\pm$ 0.10 \\
23438 & 3.69 & 967 $\pm$ 115 & 42.42 $\pm$ 0.02 & 6.70 $\pm$ 0.11 & 0.35 $\pm$ 0.09 & 9.16 $\pm$ 0.13 \\
28074 & 2.26 & 1126 $\pm$ 17 & 43.31 $\pm$ 0.01 & 7.29 $\pm$ 0.01 & 0.84 $\pm$ 0.03 & 9.54 $\pm$ 0.04 \\
28812 & 4.22 & 1009 $\pm$ 123 & 42.85 $\pm$ 0.03 & 6.95 $\pm$ 0.13 & 0.56 $\pm$ 0.25 & 8.70 $\pm$ 0.10 \\
31747 & 4.13 & 523 $\pm$ 16 & 42.45 $\pm$ 0.01 & 6.23 $\pm$ 0.03 & 1.55 $\pm$ 0.11 & 8.53 $\pm$ 0.15 \\
53501 & 3.43 & 753 $\pm$ 98 & 42.62 $\pm$ 0.03 & 6.56 $\pm$ 0.12 & 0.73 $\pm$ 0.20 & 9.15 $\pm$ 0.07 \\
73488 & 4.13 & 681 $\pm$ 64 & 42.76 $\pm$ 0.02 & 6.58 $\pm$ 0.09 & 1.21 $\pm$ 0.24 & 9.44 $\pm$ 0.07 \\
119957 & 4.15 & 1070 $\pm$ 80 & 42.45 $\pm$ 0.03 & 6.81 $\pm$ 0.07 & 0.30 $\pm$ 0.05 & 8.73 $\pm$ 0.08 \\
144195 & 3.35 & 1117 $\pm$ 81 & 42.49 $\pm$ 0.03 & 6.90 $\pm$ 0.07 & 0.31 $\pm$ 0.06 & 8.59 $\pm$ 0.09 \\
154183 & 3.55 & 570 $\pm$ 132 & 42.84 $\pm$ 0.03 & 6.49 $\pm$ 0.21 & 2.08 $\pm$ 1.07 & 8.42 $\pm$ 0.10 \\
167741 & 4.12 & 1348 $\pm$ 146 & 42.55 $\pm$ 0.02 & 6.99 $\pm$ 0.10 & 0.18 $\pm$ 0.05 & 9.53 $\pm$ 0.18 \\
\enddata
\tablecomments{Column (1): srcid number, same as \cite{Anna2025c} catalog. Columns (2): redshifts derived from the NIRSpec spectra. Column (3)-(4): The FWHM (after the instrumental broadening correction) and luminosity of the broad H$\alpha$ emission line components. Column (5): The log scale blackhole masses derived from the H$\alpha$ emission line. Column (6): The Eddington ratio. Column (7): The stellar mass estimation results derived from the UV luminosity.}
\end{deluxetable*}

\begin{deluxetable}{lccc}
\tablecaption{\centering The properties of H$\beta$ and Fe~\textsc{II} in our sample
\label{tab: info_hb}}
\tabletypesize{\footnotesize}
\tablehead{
\colhead{srcid} & \colhead{$\rm FWHM_{\rm H\beta}$} & \colhead{$\log L_{\rm H\beta,~broad}$} & \colhead{$\log L_{\rm Fe}$} \\
 & \colhead{$\rm km~s^{-1}$} & $\log(\rm erg~s^{-1})$ & $\log(\rm erg~s^{-1})$ \\
\colhead{(1)} & \colhead{(2)} & \colhead{(3)} & \colhead{(4)} 
}
\startdata
28074 & 1439 $\pm$ 112 & 41.89 $\pm$ 0.03 & $<42.52$ \\
53501 & 569 $\pm$ 120 & 41.32 $\pm$ 0.08 & $<42.55$ \\
73488 & 1049 $\pm$ 221 & 41.14 $\pm$ 0.08 & $<42.42$ \\
\enddata
\tablecomments{Column (1): srcid number, same as \cite{Anna2025c} catalog. Columns Column (2)-(3): The FWHM and luminosity of the broad H$\beta$ emission line components. Column (4): The upper limit of total luminosity of Fe lines in $4434–4684\rm \AA$}
\end{deluxetable}
  
\section{Results} \label{sec: results}
In this section, we present the correlations between different emission-line components and continuum luminosities, estimate the black hole properties of our LRD sample, and compare these properties with those of normal AGNs.

\subsection{Relation Between Emission Line and Continuum Luminosity}
\begin{figure*}
\includegraphics[width=1.0\textwidth]{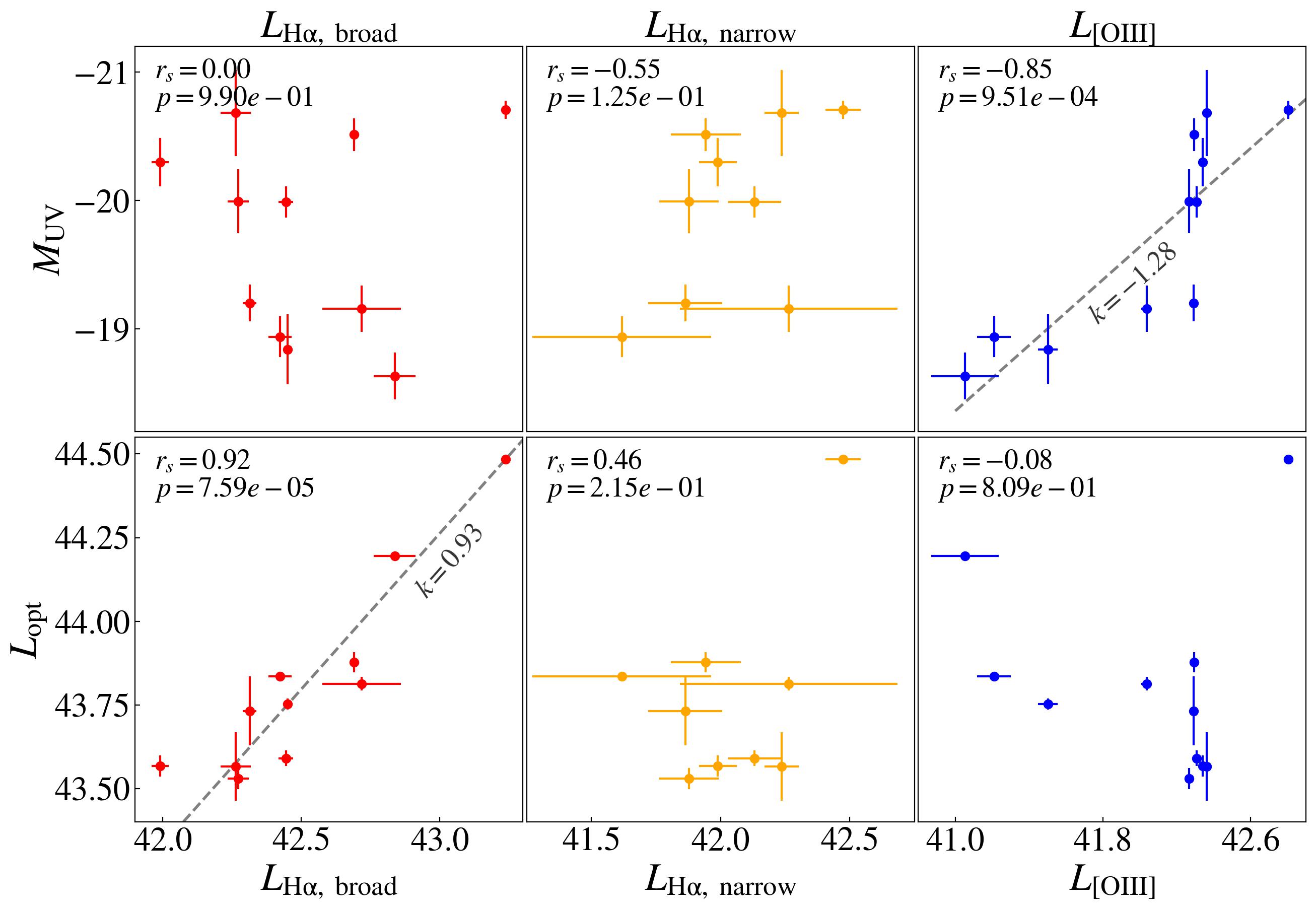}
\caption{The flux of the broad H$\alpha$ component, the narrow H$\alpha$ component, and the [O~\textsc{III}]$\lambda5007$ emission line as a function of rest-frame UV and optical luminosities from \cite{Anna2025c} in our LRD sample. Spearman correlation coefficients and p-values are indicated in each panel. The broad H$\alpha$ flux shows a strong correlation with the optical luminosity but exhibits no significant correlation with the UV magnitude. Conversely, the [O~\textsc{III}] flux correlates with the UV magnitude rather than the optical luminosity. The slopes of significant correlation are noted in the corresponding sub-panels. The narrow H$\alpha$ component displays weak correlations with both the UV and optical bands.
\label{fig: correlation}}
\end{figure*}

Figure \ref{fig: correlation} illustrates the relationship between the broad and narrow H$\alpha$ component fluxes, as well as the [O~\textsc{III}]$\lambda5007$ flux, and the absolute UV magnitude ($M_{\rm UV}$) at $1500~\rm \AA$ and the optical continuum luminosity. Spearman correlation coefficients and corresponding p-values are indicated in the upper right corner of each panel. It's worth noting that only 4 of 11 sources have M/H observations of the [O~\textsc{III}] lines; the measurements for the remaining sources are from \cite{Anna2025c}.

As shown in the first column, the flux of the broad H$\alpha$ component exhibits a strong correlation with the optical luminosity but shows virtually no correlation with the UV magnitude. A natural explanation for this result is that the broad H$\alpha$ emission and the optical continuum share a common origin, both likely powered by the central black hole accretion process. In comparison, \cite{Anna2025c} also reports a tight correlation between the total H$\alpha$ luminosity and the rest-frame optical continuum, which is partly consistent with our findings, given that the broad component dominates the H$\alpha$ flux in most of our LRD sample (The median flux fraction of the broad component is 74\%). A more detailed analysis of the H$\alpha$ emission-line components was recently presented by \cite{Asada26}, who inferred that both the broad and narrow Gaussian components of H$\alpha$ correlate with 
UV luminosity. However, their study primarily used lower-resolution prism spectra, and therefore their definitions for decomposing broad and narrow components differ from ours. They also noted that the correlation for the broad H$\alpha$ components shows slightly larger scatter compared to that of the narrow components.

The second column of Figure \ref{fig: correlation} presents the results for the narrow component of H$\alpha$. Unlike the broad component, the narrow component shows only weak correlations with both UV and optical luminosity. This could be due to the lower SNR of several detections or to a more complex physical origin for the narrow H$\alpha$ emission. Statistically, the narrow H$\alpha$ component appears more likely to correlate with the UV magnitude—a trend similar to that seen for [O~\textsc{III}], described below. Nevertheless, a portion of the narrow H$\alpha$ emission may still be linked to the optical continuum, which could originate from the thinner gas outside the gas shell structures in LRDs, an analogy of the narrow-line region (NLR) in AGN.

The last column of Figure \ref{fig: correlation} shows that the [O~\textsc{III}] flux strongly correlates with the UV magnitude but shows almost no correlation with the optical luminosity. This suggests that the far-UV photons extended from the UV continuum are the most likely source of ionizing photons for the [O~\textsc{III}]  lines, which is consistent with \cite{Anna2025c} and \cite{Asada26}, although the UV magnitude may have relatively large uncertainty due to the limited fitting wavelength in the rest-frame prism spectra. The origin of both the UV continuum and the [O~\textsc{III}] emission still remains difficult to constrain with current observations. In general, three main possibilities are considered:

\begin{itemize}
    \item [1.] {\bf Stellar Origin}. One explanation attributes the UV continuum to recently formed stars in the host galaxy. Observationally, this could account for the extended structures seen in the rest-frame UV for some LRDs. Furthermore, extreme starburst galaxies \citep[e.g., the extreme emission-line galaxies in JADES,][]{2024MNRAS.535.1796B} can produce the high [O~\textsc{III}] EWs measured in LRDs \citep{Anna2025c, 2025arXiv251215853B}. This origin is also suggested by the empirical decomposition process of prism spectra of LRD \citep{sun2026littlereddot}.
    \item [2.] {\bf AGN Origin}. Another interpretation is that both the [O~\textsc{III}] emission and the UV continuum originate from the narrow-line region and the inner accretion disk of a central AGN. However, compared to the standard AGN unification model \citep{1993ARA&A..31..473A,2015ARA&A..53..365N}, two issues need to be addressed. First is the large EWs of the [O~\textsc{III}] line. Even when calculated using the observed (relatively red) LRD continuum (which already overestimates the available ionizing photons), the [O~\textsc{III}] EWs are higher than those in most type-I \citep{2011ApJS..194...45S} or type-II \citep{2010A&A...510A..56B,2011MNRAS.415.1928C} AGN. This may require a larger covering factor of the NLR, specific gas-phase metallically, or lower gas densities. Second is the viewing-angle constraint: if AGN-originated UV photons can escape along certain lines of sight and excite more distant narrow-line gas, we would expect to observe a population of “bluer dots” with UV luminosities significantly exceeding those from star formation and showing shorter UV variability timescales. To date, no such population has been statistically confirmed though several cases have been observed \citep[e.g.,][]{2026arXiv260122214B}.
    \item [3.] {\bf Diffuse Ionized Gas Origin}. This scenario suggests that the UV photons come from recent star formation in the host galaxy, while the [O~\textsc{III}] emission is collisionally excited by more tenuous gas located outside the central black hole’s gas shell. This region more closely resembles the photo-ionization conditions of diffuse ionized gas (DIG) rather than those of classical H~\textsc{II} regions in the local universe. This may cause a mild FWHM in [O~\textsc{III}] line, which relates to the stellar feedback and gas removal around the central accreting BH \citep{Kohei2025model}.    
\end{itemize}

\subsection{Location and Evolution of LRD in AGN diagram}
We estimate the black hole mass ($M_{\rm BH}$), bolometric luminosity ($L_{\rm bol}$), and Eddington ratio ($\lambda_{\rm Edd}$) for our sample based on the broad H$\alpha$ emission-line features. Systematic uncertainties in the $M_{\rm BH}$ estimation, including contributions from electron scattering, are discussed in Section \ref{subsec: mbh_disuss}. Specifically, The $M_{\rm BH}$ is derived using the virial mass estimator calibrated by \cite{2013ApJ...775..116R}:

\begin{equation} \label{formula1}
\begin{split}
    \mathrm{l}&\mathrm{og}(M_{\mathrm{BH}}/M_{\odot})= \log(\epsilon) + 6.57 
    \\
    &+ 0.47\log \left(\frac{L_{\mathrm{H}\alpha,~\mathrm{broad}}}{10^{42} \mathrm{~erg~s^{-1}}} \right) + 2.06\log\left(\frac{\mathrm{FWHM}_{\mathrm{H}\alpha}}{\mathrm{10^{3}~km~s^{-1}}}\right)^2
\end{split}
\end{equation}

where $L_{\rm H\alpha,~broad}$ is the luminosity of the broad H$\alpha$ component and $\rm FWHM_{\rm H\alpha,~broad}$ is its full width at half maximum after correcting for instrumental broadening. A scaling factor of $\epsilon=1.075$ is adopted following \cite{2015ApJ...813...82R}. The instrumental broadening correction was applied using the wavelength-dependent resolution of NIRSpec, with revisions on the resolution of medium/high-resolution spectra \citep{2025A&A...702L..12S}. Note that we do not correct $L_{\rm H\alpha,~broad}$ for extinction. The bolometric luminosity is estimated using the H$\alpha$-based calibration $L_{\rm bol}=130\times L_{\rm H\alpha,~broad}$ from \cite{2012MNRAS.423..600S} (see more discussion about the uncertain of using such relation to LRDs in Section \ref{subsec: lbol_disuss}). The Eddington ratio is then calculated as $\lambda_{\rm Edd}=L_{\rm bol}/L_{\rm Edd}$ with $L_{\rm Edd}=1.26\times 10^{38}(M_{\rm BH}/M_{\odot})~\rm erg~s^{-1}$ \citep{1986rpa..book.....R, 2011ApJS..194...45S}. The estimated $M_{\rm BH}$ and $\lambda_{\rm Edd}$ results are compiled in Table \ref{tab: info_ha}. 

\begin{figure*}
\begin{center}
\includegraphics[width=0.7\textwidth]{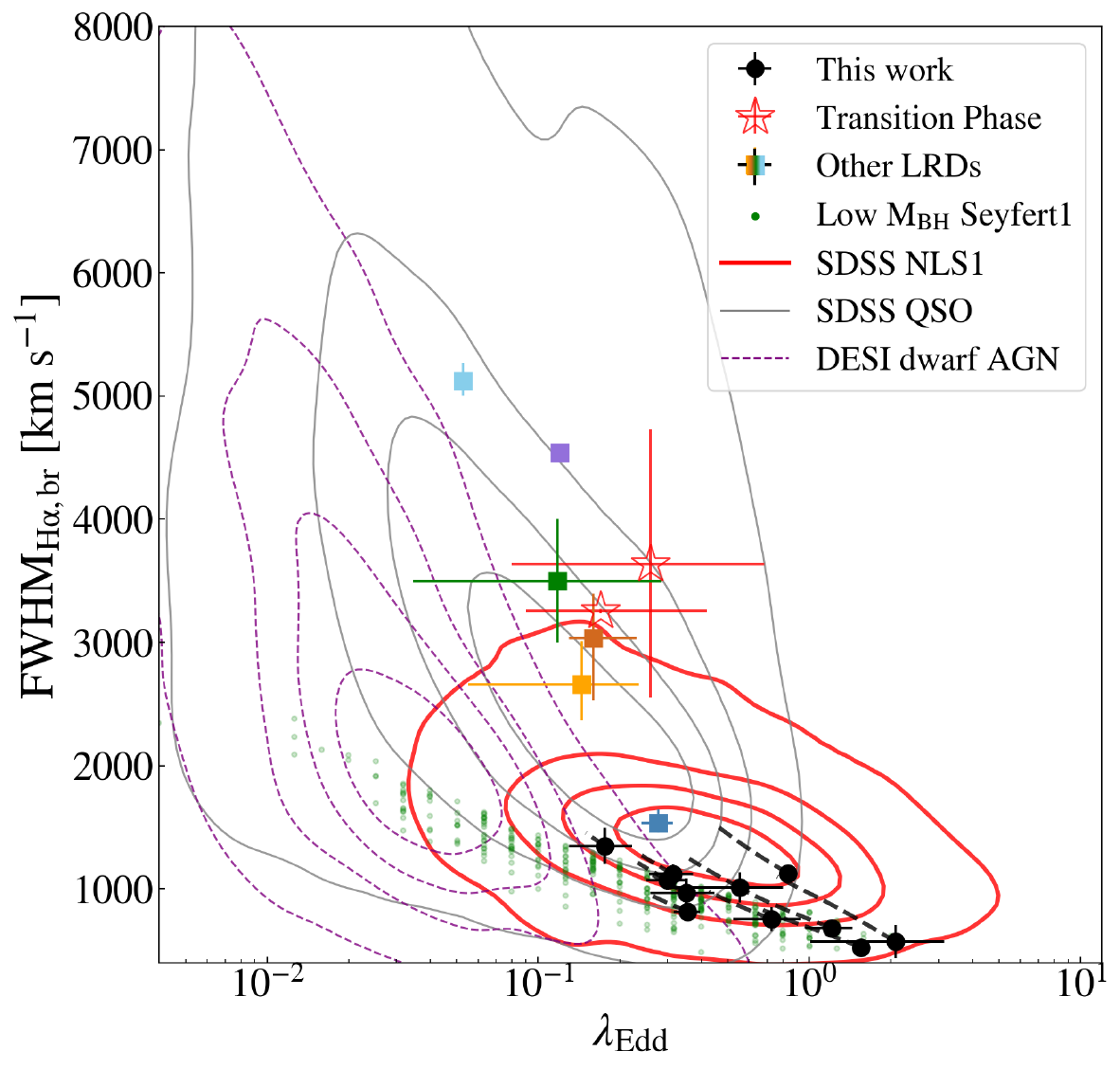}
\end{center}
\caption{Distribution of LRDs and AGNs on the $\rm FWHM_{H\alpha}–\lambda_{Edd}$ plane. Our LRD sample is shown as black filled circles. Other known LRDs from \cite{2024arXiv241204557L,2025A&A...701A.168D,2025arXiv250316596N,2025ApJ...984..121W,2025ApJ...989L...7T,2025MNRAS.544.3900J} are indicated by colored squares (using the same method from H$\alpha$ line in this work to estimate the $\lambda_{Edd}$). Recently identified candidates in a possible transitional phase of LRDs from \cite{2025arXiv251202093L,2025arXiv251202096F} are represented by open red stars. For comparison, the black, purple, and red contours show the distributions of SDSS quasars (QSOs) from \cite{2022ApJS..263...42W}, DESI dwarf AGNs from \cite{2025ApJ...982...10P}, and narrow-line Seyfert 1 (NLS1) galaxies from \cite{2024MNRAS.527.7055P}, respectively (the contour lines contains 95\%, 71\%, 48\%, and 23\% sources). Seyfert 1 galaxies with low-mass black holes from \cite{2012ApJ...755..167D} are also included as green dots (the vertical pattern is due to the $0.1~\rm dex$ precision in $\lambda_{\rm Edd}$ parameter). The black dashed lines trace the projected evolution over the next 20 million years for our LRD sources, assuming the mass accretion rate remains roughly constant.
\label{fig: FWHM_Edd}}
\end{figure*}

The distribution of LRDs and other AGN populations on the $\rm FWHM_{H\alpha}–\lambda_{Edd}$ plane is shown in Figure \ref{fig: FWHM_Edd}. Despite potential systematic uncertainties in the mass estimation, the majority of the LRD sample has inferred black hole masses of $10^6\sim 10^8 M_{\odot}$ and exhibits relatively high Eddington ratios between 0.3 and 2. These properties place them in an extreme early evolutionary phase in the AGN main sequence
, characterized by lower black hole masses compared to typical narrow-line Seyfert~1 (NLS1) galaxies \citep{2024MNRAS.527.7055P} and higher accretion rates compared to DESI dwarf AGNs \citep{2025ApJ...982...10P}. Assuming a slim-disk accretion model \citep{2000PASJ...52..133W}, the implied current mass accretion rate is $0.03\sim1~M_{\odot}~\rm yr^{-1}$. Following the model by \cite{Kohei2025model}, which derived a relatively stable mass accretion rate, we can further infer the history and evolutionary pathway of LRDs. If the LRDs are seeded by low-mass black holes ($M_{\rm BH,~seed}\ll10^6~M_{\odot}$), then the estimated age of the LRD phase is about $10^5-10^7$ years, consistent with the estimation results of the ``quasistar" evolutionary model prediction \citep{2026ApJ...996...48B,Kohei2025model}.

The dashed lines in Figure \ref{fig: FWHM_Edd} trace the evolution of LRDs over an additional $20~\rm Myr$, a typical LRD lifetime in the \cite{Kohei2025model} model, assuming the current mass accretion rate is maintained. Such a timescale is similar to recent estimation from the prism stacking spectra of LRD \citep{2026arXiv260220247P}. The endpoints of these tracks indicate that all of our LRDs would settle within the parameter space occupied by NLS1 galaxies. This suggests that, even after transitioning into an active AGN phase, these systems would still have an evolutionary path ahead before reaching the properties of typical QSOs observed in SDSS \citep[e.g.,][]{2022ApJS..263...42W}. This scenario is also consistent with the recent proposed ``transitional phase" of LRDs, characterized by a similar ``V-shaped" SED but with clear X-ray detections \citep{2025arXiv251202093L,2025arXiv251202096F}, which are located at the edges of NLS1 samples. 

In Figure \ref{fig: FWHM_Edd}, we also plot the locations of the most extreme LRDs identified to date, although their properties are mainly based on prism spectra \citep{2024arXiv241204557L,2025A&A...701A.168D,2025arXiv250316596N,2025ApJ...984..121W,2025ApJ...989L...7T,2025MNRAS.544.3900J}. These extreme LRDs occupy a region distinct from that of our sample, which can be explained by two possibilities. Their emission-line widths could be affected by strong electron scattering, which would place them in a region corresponding to lower black hole masses and higher accretion rates. Alternatively, the formation history and evolution of these extreme LRDs may be governed by different initial conditions.


\subsection{The Strength of Iron Lines of LRDs}
The Fe~\textsc{II} emission lines are common and diagnostically important features in Type-I AGN spectra \citep{1977ApJ...215..733O,1985ApJ...288...94W,1992ApJS...80..109B}. In the standard AGN model, optical Fe~\textsc{II} emission is thought to originate in the outer part of the BLR \citep{2010ApJS..189...15K}. Their line shape and EW can reflect the physical conditions of the emitting clouds \citep{2004ApJ...615..610B}. Observationally, the optical Fe~\textsc{II} lines around H$\beta$ and [O~\textsc{III}] form the major part of Eigenvector1 \citep[or the quasar main sequence,][]{2004AJ....128.2603Y,2014Natur.513..210S}, exhibiting strong correlations between Fe~\textsc{II} strength, [O~\textsc{III}] EW, and other optical spectral properties. The Fe~\textsc{II} strength, quantified by the flux ratio $R_{\rm FeII} = F_{\rm Fe}/F_{\rm H\beta}$, is an important diagnostic for the Eddington ratio and black hole mass scaling relations. Recently, \cite{2025arXiv251216981T} reported weak Fe II emission, possibly a sign of metal-poor BLRs in high-redshift QSOs.

\begin{figure}
\includegraphics[width=0.48\textwidth]{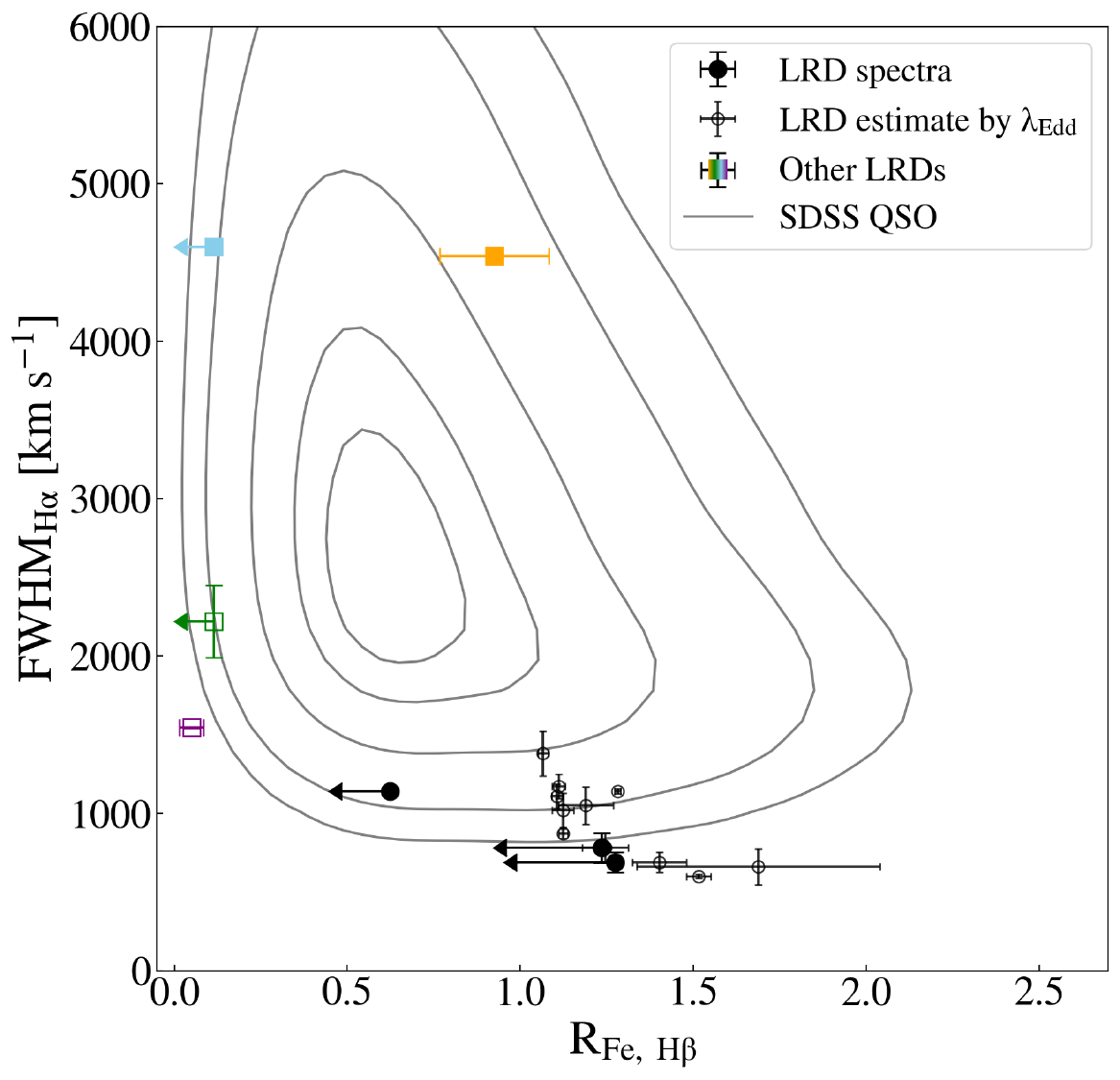}
\caption{Distribution of LRDs and AGNs on the $\rm FWHM_{H\alpha}–R_{FeII}$ plane. Upper limits for our LRD sample are plotted as black filled dots, while predicted $\rm R_{FeII}$ values based on the $\rm \lambda_{Edd}-R_{FeII}$ relation from the QSO sample in \cite{2025ApJ...987...48P} are indicated by open black dots. Results from other literature studies with Fe line fitting are summarized as squares \citep{2024arXiv241204557L,2025arXiv251000103T,2025ApJ...994L...6T,2025arXiv251000101D}. The permitted and forbidden line detection results are represented by solid and open squares, respectively.
The background gray contours represent the distribution of SDSS QSOs from \cite{2022ApJS..263...42W}. Here, $\rm R_{FeII}$ is calculated using the Fe flux within the rest-frame wavelength range of $4434–4684\rm \AA$.
\label{fig: FWHM_Fe}}
\end{figure}

In our sample, three sources have detected broad H$\beta$ emission, but none show a significant detection of the associated Fe~\textsc{II} lines. Adopting the $R_{\rm FeII}$ definition which estimate the Fe flux spanning from 4434 to 4684$\rm \AA$ from \cite{2022ApJS..263...42W} and \cite{2025ApJ...987...48P}, we estimate 3$\sigma$ upper limits for $R_{\rm FeII}$, shown as black filled circles in Figure \ref{fig: FWHM_Fe}. For comparison, the open circles represent the predicted $R_{\rm FeII}$ values inferred from the $\rm \lambda_{Edd}-R_{FeII}$ relation from the QSO sample from DESI QSOs by \cite{2025ApJ...987...48P}. Our upper limits lie systematically below these predictions. We further compile recent observational constraints on Fe lines in LRDs: \cite{2024arXiv241204557L} report detections of broad Fe~\textsc{II} lines, with line detection in 4434 to 4684$\rm \AA$, their LRD is unprecedented luminous; \cite{2025ApJ...994L...6T} detect four broad optical Fe~\textsc{II} lines in four sources, though not within the $R_{\rm FeII}$ calculation range; \cite{2025arXiv251000101D} and \cite{2025arXiv251000103T} detect narrower forbidden [Fe~\textsc{II}] lines. Compared to the distribution of SDSS QSOs, the measured (or constrained) Fe~\textsc{II} values in LRDs are generally lower than expected given their high Eddington ratios, indicating relatively weak permitted Fe~\textsc{II} emission in these systems.

Furthermore, the forbidden [Fe~\textsc{II}]/[O~\textsc{I}] line ratios presented by \cite{2025arXiv251000101D} suggest that the metallicity of the Fe-emitting gas is not particularly low. This argues against the possibility that the weakness of the permitted Fe~\textsc{II} lines is primarily due to low metallicity, although we note that forbidden and permitted Fe~\textsc{II} lines trace different gas phases \citep{2025arXiv251000103T}. Another relevant clue from \cite{2025ApJ...994L...6T} is that the permitted Fe~\textsc{II} lines in their LRDs have lower FWHM than the Balmer lines, implying an origin at larger radii. This stratified structure is reminiscent of that observed in normal AGN. We therefore suggest that the weakness of the Fe~\textsc{II} emission in LRDs is likely caused by a deficit of ionizing photons in the outer regions of the gas shell. We will discuss more details of this scenario in Section \ref{subsec: gasshell}.

\section{Discussion} \label{sec: discuss}
In this section, we first discuss the systematic uncertainties in the estimates of $M_{\rm BH}$ and $L_{\rm bol}$. Then we investigate the characteristic size, mass, and structure of the LRD gas shell, and finally propose a model with a thickened photosphere, analyzing its implications for the observed optical continuum under different physical conditions. Finally, we examine the possible $M_{\rm BH}$–$M_{*}$ relation in our sample.

\subsection{The uncertainty of FWHM and $M_{\rm BH}$ estimation} \label{subsec: mbh_disuss}

Recent work by \cite{2025arXiv250316595R} and \cite{2026arXiv260118864S} suggests that electron scattering can significantly broaden the Balmer emission lines. A key piece of evidence is that for high SNR H$\alpha$ line profiles, models that convolve a single Gaussian with an exponential wing could yield the best fit. Specifically, two sources in our sample overlap with those analyzed by \cite{2025arXiv250316595R}: srcid-73488 (our $\log(M_{\rm BH}/M_{\odot}) = 6.58$ versus their value of 6.1) and srcid-53501 (our $\log(M_{\rm BH}/M_{\odot}) = 6.55$ versus their upper limit of 5.6). 

Specifically, the difference in black hole mass estimates primarily originates from the intrinsic FWHM derived from the line fitting. Our FWHM measurements are based on the composite line profile obtained by summing the two broad Gaussian components, which essentially describes the emission-line profile excluding the narrow-line core. To account for the effect of an exponential profile on the intrinsic line width, we fitted the H$\alpha$ lines in our sample using a model consisting of a broad Gaussian component with its electron scattering component, which is represented by convolving with an exponential profile, following the model of \cite{2025arXiv250316595R} and \cite{matthee2026engineflowslittlered}. In terms of the Bayesian Information Criterion (BIC), the two-Gaussian and exponential models yield generally comparable results for most sources: six sources have BIC differences smaller than 10, and three sources show a preference for the double-Gaussian model. Only srcid-73488 and srcid-167741 exhibit a slight preference for the exponential model, with BIC differences of approximately 10–15. This indicates that both models provide reasonable fits. The minor discrepancy with the results of \cite{2025arXiv250316595R} can be attributed to the fact that our two-Gaussian fit does not fix the central wavelengths, allowing for a better representation of asymmetric line wings.

We also compared the FWHM of the total line profile derived from the two-Gaussian model with that from the broad Gaussian kernel in the exponential model. The overall systematic scatter between the two is about 0.11 dex, and the FWHM from the exponential model is, on average, 16\% lower than that from the two-Gaussian model. We consider these results to provide a reasonable estimate of the systematics introduced by electron scattering.

Finally, the term $L_{\rm H \alpha}$ in Eq.~\ref{formula1} is based on the $L_{5100}-L_{\rm H \alpha}$ relation. To assess the impact of this dependence, we compare the such relation in our sample using $L_{\rm H \alpha, broad}$ from our fitting results and $L_{5100}$ from \cite{Anna2025c}  with the relation from \citep{2013ApJ...775..116R}. We find that the deviation of $L_{5100}-L_{\rm H \alpha}$ is about 0.23dex, which is similar to the EWs difference between our sample (average 429$\AA$, median 420$\AA$) from SDSS DR7 QSOs \citep[average 282$\AA$, median 259$\AA$,][]{2011ApJS..194...45S}. Such a difference may lead to an overestimate of $M_{\rm BH}$ by about 30\%.

A lower black hole mass implies a correspondingly higher Eddington ratio for these LRDs. Even without correcting the optical continuum luminosity for dust extinction, the inferred luminosities are already comparable to $L_{\rm Edd}$. Given the relatively small values of $M_{\rm BH}$, this implies black hole growth timescales of $\lesssim 10^4$–$10^5$ yr, as well as short gas-shell depletion timescales of order $10^3$–$10^4$ yr \citep{2026arXiv260118864S}.
A promising method to distinguish between these models is to study the variability timescale of the H$\alpha$ line, particularly its wings. The electron scattering scenario predicts a smaller BLR size, and the associated ionization structure would respond more rapidly to changes in the ionizing continuum than the dynamical timescale of the BLR \citep{2006ApJ...643..112L}.

It is important to acknowledge that single-epoch spectroscopic mass estimates carry an intrinsic systematic uncertainty 
due to uncertainties in the virial factor and other BLR structural parameters \citep{2011nlsg.confE..32P,2012ApJ...753..125S,2018ApJ...856....6D}. A dedicated reverberation-mapping campaign for LRD objects is crucial to provide an empirical calibration for black hole mass estimates in this population.


\subsection{The uncertainty of  $L_{\rm bol}$ estimation} \label{subsec: lbol_disuss}


Since our sample lacks direct photometric coverage from X-ray to far-infrared, we have to rely on bolometric corrections (BC). BC based on the AGN continuum (e.g., at 5100$\AA$) assumes that the emission originates from the central accretion disk \citep{2006ApJS..166..470R,2012MNRAS.422..478R}, which is unlikely to hold for LRDs. On the other hand, the H$\alpha$-based BC considers the connection between H$\alpha$ and far-UV ionizing photons \citep{1995ApJ...455L.119B}. The underlying physical mechanism is thus similar in AGN and LRDs, although the exact calibration factor could be changed because of the difference in the covering factor and the central engine structure. We have compared $L_{\rm bol}$ derived from H$\alpha$ with that estimated from $5100\rm \AA$ continuum using the bolometric correction from \cite{2006ApJS..166..470R}. The two estimates are consistent within a scatter of $0.2~\rm dex$. Following \cite{2026ApJ...996..129G}, who find that $L_{\rm bol}/L_{\rm 5100}$ may be overestimated by a factor of 2, combined with the $L_{5100}-L_{\rm H \alpha}$ deviation estimated in Section \ref{subsec: mbh_disuss}, we estimate that our $L_{\rm bol}$ may be overestimated by a factor of about 3.

Overall, considering the differences in structure and SED between LRDs and normal AGN, the $M_{\rm BH}$ may be overestimated by roughly a factor of two, while $L_{\rm bol}$ could be overestimated by a factor of three. Since both quantities relate to the deviation of the $L_{\rm H\alpha}$ relation, the resulting $\lambda_{\rm Edd}$ uncertainty is about a factor of two.


\subsection{Possible Clumpy Gas Envelope of LRDs} \label{subsec: gasshell}

\begin{figure*}
\includegraphics[width=1\textwidth]{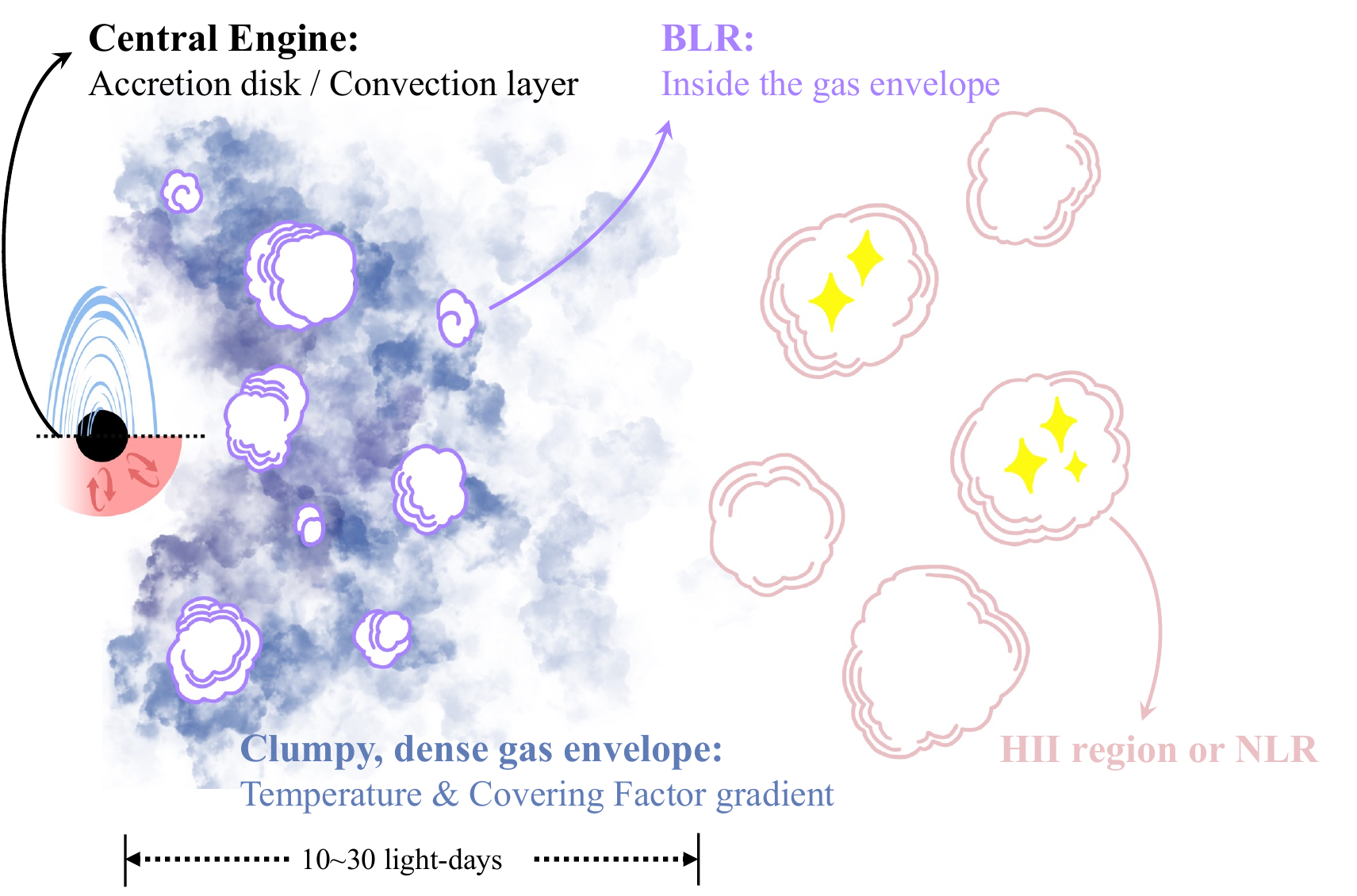}
\caption{Schematic illustration of our proposed gas shell structure for LRDs. Compared to the cocoon envelope model originally proposed by \cite{2025arXiv250316596N}, our model suggests a larger physical size and lower covering factor in each layer of the cocoon envelope. Compared to the black hole star model of \cite{Kohei2025model}, our model possesses a thicker photosphere, with a temperature that gradually decreases from the inner to outer regions, resulting in a larger outer radius for the gas shell. Relative to the model presented by \cite{2025arXiv250710659L}, our model is also compatible with an asymmetrical BLR structure.
\label{fig: LRD_model}}
\end{figure*}

Within the framework where the optical emission originates from the central black hole accretion, two main models currently seek to explain the origin of the optical continuum and broad emission lines in LRDs. The first posits a supermassive black hole accreting from an asymmetric, dense gas cocoon, where the inner region is ionized and surrounded by a much larger reservoir of cold gas \citep[e.g.,][]{2025arXiv250316595R,2026arXiv260118864S}. In this picture, the broad emission lines could be produced by electron scattering in the inner region, while the continuum is shaped by the outer gas. The second model analogizes the LRD's core to a ``quasistar", where the continuum arises from a stellar-like photosphere, and the broad-line region (BLR) resides inside or outside this photosphere \citep[e.g.,][]{2025MNRAS.544.3407K,Kohei2025model}. In this section, we propose a model revision featuring a thicker, clumpy gas envelope. This structure can produce blackbody-like emission at larger radii while also accommodating the ionized clouds responsible for broad emission lines.

As presented in Section \ref{sec: results}, the strong correlation between the broad-line luminosity and the optical continuum luminosity strongly suggests a common origin. Moreover, the large Balmer decrement observed in LRD \citep[$>10$ for both broad H$\alpha$ and H$\beta$ detected in our sample; see also][]{2025ApJ...986..177B} indicates that the broad lines likely form within a dense gas with $n_{\rm H}\geq10^8\rm~cm^{-3}$ \citep{2025arXiv251211050Y}, which could also account for the Balmer absorption features seen in a proportional of LRDs \citep[e.g.][]{2024ApJ...974..147L,2024MNRAS.535..853J}. The most natural interpretation of these observations is that the BLR is embedded within the dense gas shell; equivalently, the outer radius of the shell (i.e., the photosphere) must exceed the characteristic size of the BLR.

We first estimate the BLR size in LRDs to constrain the minimum outer radius of the gas shell. Using the established $R-L$ relation for traditional AGNs, we estimate the radius of the H$\alpha$-emitting BLR clouds to be 22–42 light-days for our sample (spanning the 17th to 84th percentiles, hereinafter the same). On the other hand, we can estimate the photospheric radius by comparing the intrinsic blackbody emission to the observed optical luminosity. Using the continuum fitting results from \cite{Anna2025c}, we estimate a radius of 6–11 light-days for a single-temperature blackbody with a 100\% covering factor. This value is smaller than the BLR size estimated by the $R-L$ relation, presenting a potential contradiction.

This tension could be resolved in different ways, depending on the specific properties of LRDs, as it involves the $R-L_{\rm ion}$ ($L_{\rm ion}$ for the ionizing luminosity), $L_{\rm ion}-L_{\rm 5100}$, and $L_{\rm 5100}-L_{\rm H\alpha}$ relations \citep{2019ApJ...886...42D}. For some extreme LRDs with very high H$\alpha$ EW, such tension could be explained by an unusual 
$L_{\rm 5100}-L_{\rm H\alpha}$ relation, which represents an extremely high covering factor for the H$\alpha$-emitting clouds compared to $\sim$20\% in normal AGNs \citep{1995ApJ...455L.119B,2022FrASS...950409P}. However, since our estimates in Section \ref{subsec: mbh_disuss} do not show a large deviation, alternative explanations are: (1) the ionizing source is not isotropic or is not a point source, making a larger uncertainty on the $R-L_{\rm ion}$ relation; and (2) the SED is very different from that of normal AGN, causing a larger scatter on the $L_{\rm ion}-L_{\rm 5100}$ relation. Indeed, the recent non‑spherical cocoon model could ease the tension via the former explanation \citep{2026arXiv260118864S}, while the extremely super‑Eddington black hole star model could mitigate it via the latter \citep{Kohei2025model}.




On the other hand, when we reconsider the possible transition process of LRDs into the normal AGN phase, another possibility emerges: the blackbody‑emitting clouds have a total covering factor of 100\%, but do not occur in a single layer. In that case, the physical size of the H$\alpha$-emitting region in some LRDs could become comparable to that of AGN BLRs. Consequently, the only structural difference between LRDs and AGN might be a denser gas shell. Following this scenario, we propose a model featuring a thicker, clumpy outer gas envelope. To illustrate how this model differs from a simple stellar photosphere, we construct a toy model as follows. Consider a gas shell with an inner radius $R_{\rm in}=10$ light-days and an outer radius $R_{\rm out}=30$ light-days ($2000-5000~\rm AU$). To prevent the escape of interior photons, we assume the total line-of-sight covering factor through the entire shell is 100\%, but the covering factor at any given radial layer is much less than unity. We assume the physical cross-sectional area of optically thick clumps is constant with radius, implying that the local covering factor scales as $f(r)\propto r^{-2}$. Consequently, the luminosity incident on each layer also scales as $L(r)\propto r^{-2}$. Assuming all incident radiation is thermally reprocessed and emitted as a blackbody, with a constant emitting area in each layer, the effective blackbody temperature scales as $T(r)\propto r^{-0.5}$. From the ideal gas law $p=nkT$, where the radiation pressure $p(r)\propto L(r) \propto r^{-2}$ and $k$ is the Boltzmann constant, the gas density then scales as $n(r)\propto r^{-1.5}$. We also account for the effect of gas motion, though its influence on the continuum shape is minor. Specifically, we assume that at a radius of 13 light-days, the gas layer co-rotates with the BLR gas at a velocity of 
$1000~\rm km~s^{-1}$ (analogous to the case of srcid-23438), while at other radii within the envelope, the velocity follows a profile $v(r)\propto r^{-0.5}$. These velocity profiles are then convolved, as Gaussian broadening kernels, with the intrinsic blackbody emission at each layer.

\begin{figure}
\includegraphics[width=0.48\textwidth]{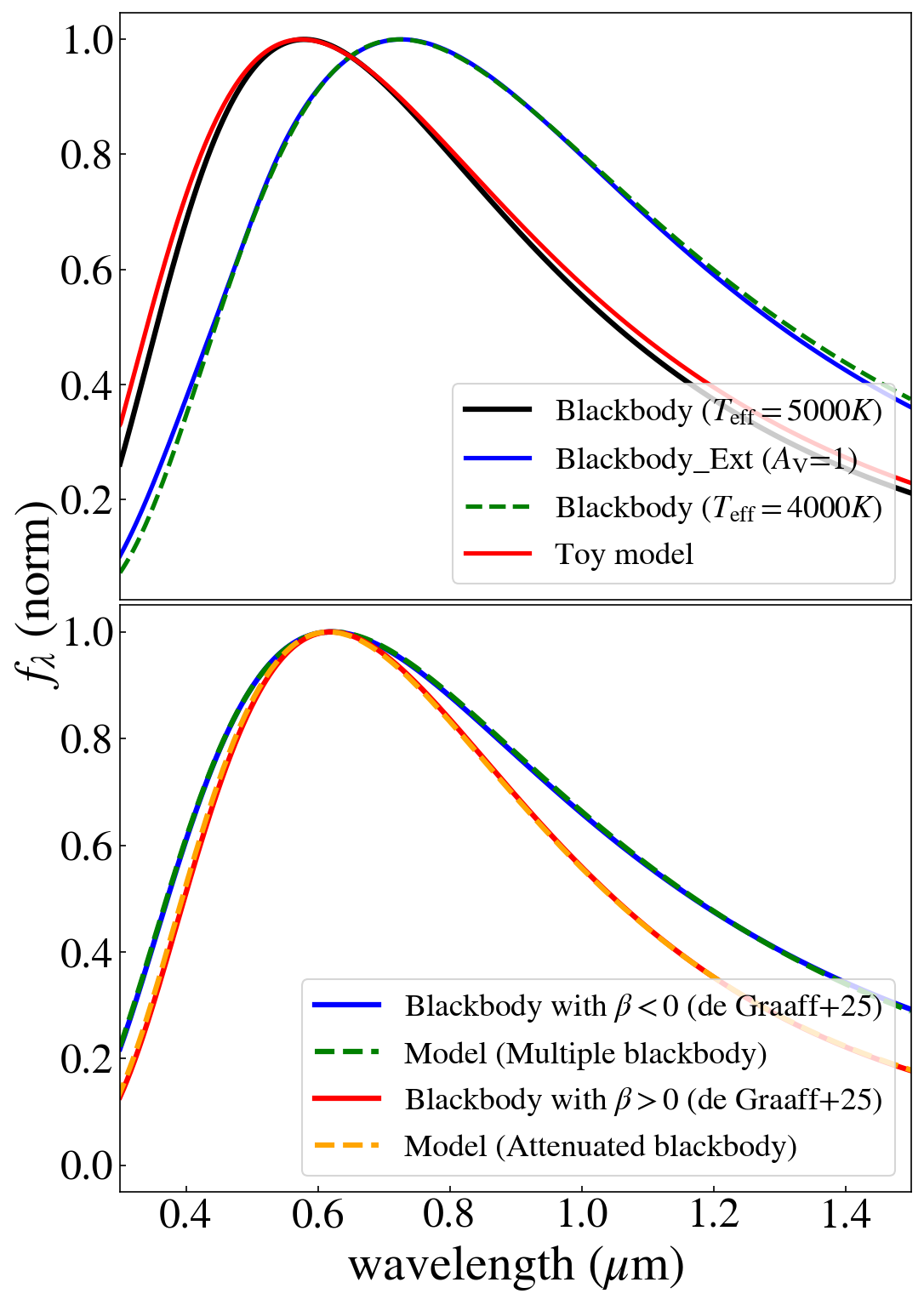}
\caption{
Physical origins and interpretations of the LRD optical continuum shape.
\textbf{Upper panel}: Influence of extinction and gas shell thickness. The black solid curve represents the standard blackbody emission for $T_{\rm eff}=5000K$. The blue curve shows the same blackbody spectrum after applying an extinction of $A_{V}=1$ assuming the extinction law from \cite{1999PASP..111...63F}, which does not differ significantly from a cooler $4000K$ blackbody (green curve).
The red curve shows the multicolor blackbody spectrum produced by a photosphere of our toy model described in Section \ref{subsec: gasshell}, which is largely consistent with the standard $5000K$ blackbody. \textbf{Lower panel}: Examples of the actual LRD continuum shape from \cite{Anna2025c}. The blue spectrum is from NEXUS-20152, which has 
$\beta<0$ (broader than a standard blackbody) and can be well fit by a combination of multiple blackbody profiles (green dashed curve) with a higher covering factor of the outer layers compared to the toy model. The red spectrum is from 
RUBIES-UDS-31747, which has $\beta>0$ (narrower than a standard blackbody). Its shape can be explained by the blackbody emission of the inner layers, which undergoes radiative transfer through a tenuous gas in the outer layer, rather than escaping directly in the toy model.
\label{fig: conti_shape}}
\end{figure}

Under these conditions, the synthesized continuum spectrum is calculated and shown in the upper panel of Figure \ref{fig: conti_shape}. For an inner boundary temperature $T_{\rm in}=6200~\rm K$, the overall spectrum is nearly indistinguishable from a single $5000~\rm K$ blackbody. In terms of total luminosity, this gas envelope with $R_{\rm in}=10$ light-days and $R_{\rm out}=30$ light-days is equivalent to a $5000~\rm K$ blackbody photosphere with a radius of $\sim9$ light-days. For a specific layer of gas in the $4700$–$5300~\rm K$ range, the integrated covering factor is roughly 20\%. The proposed gas shell is approximately three times larger than previous estimates, leaving ample volume for the ionized BLR gas to reside. 

This model can also account for the weakness of the Fe~\textsc{II} lines in LRDs compared to normal AGN as seen in Section \ref{sec: results}. Since the Fe~\textsc{II} lines have lower FWHM and originate from the outer part of the gas shell, the integrated covering factor of the inner, optically thick gas layers is higher along their line of sight. This results in a great reduction of the ionizing photons from the central engine before they reach the Fe~\textsc{II}-emitting zone. In contrast, classical Type-I AGN lack such an extensive, obscuring gas shell. Consequently, the outer gas in standard AGN receives a much higher flux of ionizing photons, leading to the stronger Fe~\textsc{II} emission typically observed.

Recently, \cite{wang2026waterabsorptionconfirmscool} analyzed two LRDs at relatively low redshifts, detecting water absorption features around $1.4~\mu m$ at the rest-frame. They inferred the presence of a temperature gradient in the outer gas shell, decreasing from about $4000~\rm K$ to $2000~\rm K$. This finding is broadly consistent with our toy model when a lower central temperature is considered. Our model predicts a variability timescale of several tens of days for both the optical continuum and the broad emission-line profiles at the rest-frame. Because the ionizing photons from the central engine have undergone radiative transfer through the extended outer envelope, we expect little or no correlated variability between the broad H$\alpha$ emission line and the optical continuum.

In summary, by increasing the thickness of the optically emitting layers within the gas shell and introducing a radial temperature gradient, while correspondingly reducing the volume-averaged covering factor, we can reproduce both the spectral shape and total flux expected from a smaller-radius blackbody photosphere with a single effective temperature of $5000~\rm K$.
This configuration allows the BLR to embed within the gas envelope, which is the main difference from the black hole star model. 
Our model differs from the dense gas cocoon framework in both scale and physical structure. While the latter typically assumes a lower black hole mass and a more compact, high-density gas configuration, our model instead favors a gas envelope that is approximately $0.5$ dex larger in spatial extent and characterized by a lower effective covering factor.
Consequently, a key observable difference between these scenarios lies in the variability timescales of the spectral components. Nevertheless, the broad Balmer lines could still be obscured and scattered by clouds in the outer, blackbody-emitting layers, leading to a large Balmer decrement and additional broadening of its FWHM. Besides, our model remains compatible with an asymmetric BLR geometry proposed by \cite{2025arXiv250710659L}. This aspect could be tested in future studies by examining potential viewing-angle effects within a larger LRD sample. A schematic illustration of our proposed gas shell structure is shown in Figure \ref{fig: LRD_model}.

\subsection{Application to the LRD optical continuum} \label{subsec: gasshell_conti}
When observed over a sufficiently broad wavelength range, the real optical continuum of LRDs often reveals systematic deviations from a pure blackbody spectrum. These deviations can be phenomenologically modeled, as in \cite{Anna2025c}, by introducing a power-law correction term, characterized by: $$f_{\nu}\propto B_{\nu}(T)\left(\frac{\nu}{\nu_{0}} \right)^{\beta}$$ where $\nu_{0}=c/5500\rm \AA$. This results in two primary cases: $\beta>0$, where the observed continuum is narrower than a blackbody when plotted against wavelength, and $\beta<0$, where it is broader. However, the physical origin of these deviations remains poorly understood.

Applying standard extinction curves to a $\sim5000~\rm K$ blackbody does not significantly alter its fundamental shape. As illustrated in the upper panel of Figure \ref{fig: conti_shape}, a $5000~\rm K$ blackbody subjected to $A_{\rm V}=1$ extinction \citep[assuming the law from][]{1999PASP..111...63F} becomes similar with a cooler $4000~\rm K$ blackbody. Therefore, the observed deviations in LRD continua are more likely intrinsic to the gas emission process itself.

We propose that the gas envelope model introduced in the previous section can, in principle, account for both types of spectral deviation. As shown in the lower panel of Figure \ref{fig: conti_shape}:

\begin{itemize}
    \item [1.] Broader continua ($\beta<0$): If each layer of the gas envelope emits as a blackbody and the foreground extinction is negligible, the emergent spectrum is a superposition of blackbodies at different temperatures. This can produce a composite spectrum that is broader than a single-temperature blackbody when the covering factor of the outer layers exceeds the value assumed in the toy model in Section \ref{subsec: gasshell}. The example in the figure is NEXUS-20152 ($T_{\rm eff}=5119$, $\beta=-0.42$), whose continuum is well-matched by the sum of three blackbodies at $4000,5000, 6000~\rm K$.
    \item [2.] Narrower continua ($\beta>0$): 
    If the emergent radiation undergoes radiative transfer through a high-density, low-column-density outer gas layer with a wavelength-dependent optical depth $\tau$, the resulting spectrum can appear narrower than that of a standard blackbody \citep[see also][]{2025ApJ...994..113L}. An example is RUBIES-UDS-31747 ($T_{\rm eff}=3827$, $\beta=1.08$), whose shape is consistent with a $3900~\rm K$ blackbody spectrum attenuated by a foreground gas with $N_{\rm H}=10^{20}~\rm cm^{-2}$ and $n_{\rm H}=10^{9}~\rm cm^{-3}$. This scenario occurs when the blackbody emission from the inner region passes through a tenuous gas in the outer layer, rather than escaping directly as assumed in the simplified toy model of Section \ref{subsec: gasshell}.
\end{itemize}

In short, the proposed gas envelope framework can reproduce the primary observed deviations of LRD continua from simple blackbody shapes. We are currently employing spectral synthesis codes (\texttt{MAPPINGS} and \texttt{CLOUDY}) within this scenario to fit the full prism continuum shapes of LRDs and derive the corresponding physical parameters of their gaseous structures.

\subsection{Upper limit of the Stellar Mass}
Accurately estimating the stellar masses ($M_{*}$) of LRDs is still challenging. This difficulty stems from the ambiguous origin of their ``V-shaped" spectrum and the strong degeneracies in decomposing the contributions of stellar light from that of the AGN
\citep[e.g.,][]{2023ApJ...957L...7K, 2024ApJ...963..128B, 2024Natur.628...57F, 2024ApJ...969L..13W, 2025ApJ...984..121W, 2025ApJ...991...37A, 2024arXiv241204557L}.

In this work, we follow the assumption that the rest-frame UV emission is predominantly stellar in origin, while the rest-frame optical is dominated by the AGN \citep{2025ApJ...989L...7T,Anna2025c,2025arXiv251215853B}. Under this assumption, we estimate $M_{*}$ using the empirical relation between UV luminosity and stellar mass from \cite{2016ApJ...825....5S}, shown in the last column in Table \ref{tab: info_ha}. 
We note that the derived stellar masses carry a substantial uncertainty of roughly 1 dex. This uncertainty arises because ongoing star formation at the epoch of observation can outshine the integrated light from older stellar populations \citep{2024ApJ...961...73N}. Since the AGN light or nebular emission may also contribute to (or even dominate) the UV flux \citep{2025ApJ...989L..12C}, our estimates could be regarded as upper limits.

\begin{figure}[H]
\includegraphics[width=0.48\textwidth]{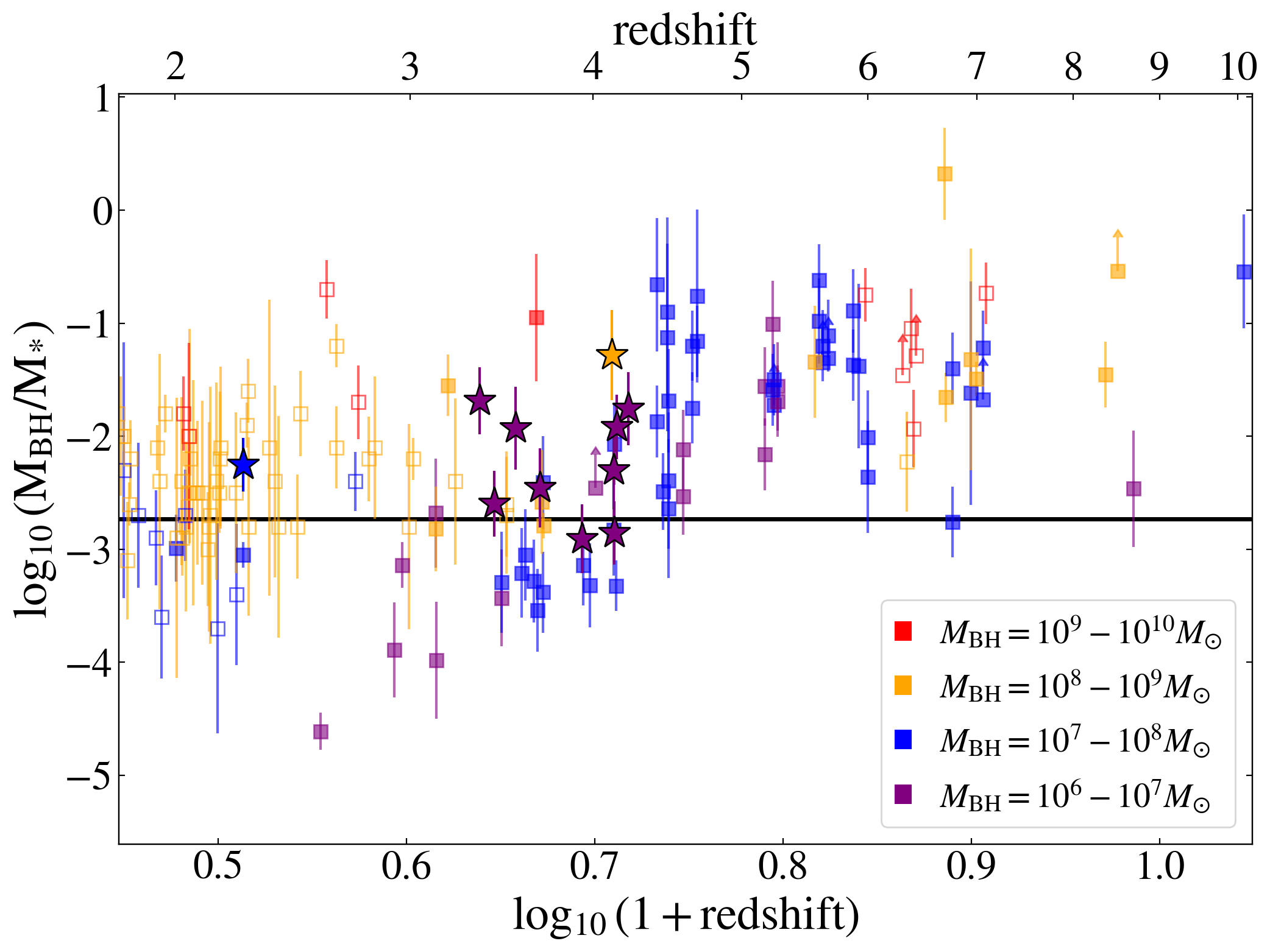}
\caption{Redshift evolution of the black hole to stellar mass ratio ($M_{\rm BH}/M_{*}$). LRD sources from this work are plotted as stars. Filled squares denote recently identified JWST selected normal AGNs and LRDs \citep{2023ApJ...955L..24G,2023ApJ...953L..29L,2023ApJ...954L...4K,2023ApJ...959...39H,2024NatAs...8..126B,2024ApJ...969L..13W,2024Natur.628...57F,2024ApJ...975..178K,2024A&A...691A.145M,2024Natur.636..594J,2025ApJ...983..165S,2025ApJ...980L..29A,2025ApJ...986..101L}. Open squares represent classical QSOs selected from ground-based surveys \citep{2023Natur.621...51D,2024ApJ...966..176Y,2024ApJ...969...11H,2025ApJ...988..234H}. All symbols are color-coded according to different black hole mass bins. As a benchmark for the local relation, the line $\log_{10} (M_{\rm BH}/M_{*})=-2.73$ (black solid line) is shown, which corresponds to the typical value of the $M_{\rm BH}/M_{\rm bulge}$ relation in the local universe for black holes in the mass range $10^6-10^{10}M_{\odot}$ \citep{2013ARA&A..51..511K}.
\label{fig: MBH_Mstar}}
\end{figure}

Using these stellar mass estimates and $M_{\rm BH}$, we compute a black hole-to-stellar mass ratio ($M_{\rm BH}/M_{*}$) of approximately 0.6\% for our sample. We present these results in Figure \ref{fig: MBH_Mstar}, where our LRDs are plotted alongside AGN samples selected both before the JWST era \citep{2023Natur.621...51D,2024ApJ...966..176Y,2024ApJ...969...11H,2025ApJ...988..234H} and from recent JWST surveys \citep{2023ApJ...955L..24G,2023ApJ...953L..29L,2023ApJ...954L...4K,2023ApJ...959...39H,2024NatAs...8..126B,2024ApJ...969L..13W,2024Natur.628...57F,2024ApJ...975..178K,2024A&A...691A.145M,2024Natur.636..594J,2025ApJ...983..165S,2025ApJ...980L..29A,2025ApJ...986..101L}.

Figure \ref{fig: MBH_Mstar} shows that our LRD sample appears slightly ``overmassive"—i.e., their $M_{\rm BH}/M_{*}$ ratios are higher—compared to the local $M_{\rm BH}-M_{\rm bulge}$
relation. We also find a tentative correlation between the degree of ``overmassive" and redshift: lower-redshift sources lie closer to the local relation, and this trend appears to hold within individual black hole mass bins, which is consistent with previous results \citep[e.g.,][]{2023ApJ...948..103Z}. 
Based on these results, we suggest a scenario in which black holes form and grow rapidly at early times, followed by the gradual buildup of stellar mass in their host galaxies, ultimately establishing the observed $M_{\rm BH}$-$M_{*}$ relation. However, this interpretation remains tentative, given the potential impact of selection biases \citep[e.g.,][]{2010ApJ...713...41S, 2023ApJ...954..173L} and the intrinsic large scatter from recent simulation results \citep[e.g.,][]{2025ApJ...984..122D}.


\section{Summary} \label{sec: summary}
In this work, we investigate the medium/high-resolution spectra of 11 LRDs at redshifts 2–4 from the sample of \cite{Anna2025c}. By decomposing the broad and narrow components of the Balmer emission lines, we analyze the correlation between the line fluxes and the continuum luminosity and estimate key physical parameters for the LRD sample. Our main conclusions are summarized as follows:

\begin{itemize}
    \item [1.] The broad component of the H$\alpha$ emission line ($\rm{FWHM}\geq600km/s$) shows a strong correlation with the optical luminosity of LRDs but exhibits no correlation with the UV magnitude. This suggests a common origin for the broad H$\alpha$ component and the optical continuum. In contrast, the [O~\textsc{III}] line strength correlates strongly with the UV magnitude but not with the optical luminosity, indicating that the [O~\textsc{III}] emission is likely excited by FUV photons from the UV continuum. However, the origin of the UV continuum and the [O~\textsc{III}] line remains unclear. The narrow component of H$\alpha$ ($\rm{FWHM}<600km/s$) displays only weak correlations with both UV and optical luminosities, implying a more complex origin.
    \item [2.] Using the FWHM and luminosity of the broad H$\alpha$ component, we estimate black hole masses in the range $10^6-10^{8}M_{\odot}$ for these LRDs, with correspondingly high Eddington ratios ($\lambda_{\rm Edd,~med}=0.6$). In the plane of the quasar main sequence, they occupy the lower-right region. Assuming a constant mass accretion rate, the growth timescale (or the age) of  LRDs in our sample is about $10^5-10^7$ year, which is consistent with the estimation results of the ``quasistar" evolutionary model. Furthermore, if the current accretion rate is sustained for an additional $\rm 20~Myr$, the predicted final state of LRDs would lie within the parameter space occupied by local narrow-line Seyfert 1 (NLS1) galaxies.
    \item [3.] From the four medium-resolution spectra covering the H$\beta$ region, we do not detect broad Fe~\textsc{II} features in the $\rm 4434–4684~\AA$ range. Combining these upper limits with a few Fe~\textsc{II} fitting attempts in other LRDs, we suggest that LRDs exhibit intrinsically weaker Fe~\textsc{II} emission compared to normal AGN. This may be attributed to structural differences of the gas shell with the AGN unified model.
    \item [4.] Using the $R-L_{\rm H\alpha}$ relation, we estimate that the BLR gas is located at distances of 10–26 light-days from the central black hole. To match the size of BLR with the luminosity from the gas shell, we propose an extended, clumpy photosphere with a temperature gradient and a smaller covering factor, rather than a thin, isothermal layer, as the source of the optical continuum. The BLR is located in the inner part of the photosphere. A simple toy model shows that the emission from such an extended gas envelope can be similar to an ideal blackbody spectrum from a much smaller radius.
    \item [5.] The diversity of the observed optical continuum shapes of LRDs (broader or narrower than a single temperature blackbody) can be explained by the thicker gas envelope as a composite of multiple blackbodies from different layers, or by self-absorption of the blackbody emission in the inner region, respectively. Future high signal-to-noise continuum observations can therefore be used to constrain the physical parameters within the photosphere.
\end{itemize}

In summary, by combining detailed emission-line information with UV and optical continuum luminosities, we have further constrained the gaseous structure and evolutionary pathway of LRDs. Our results demonstrate the crucial role of JWST NIRSpec medium- and high-resolution spectroscopy in probing the central engines of LRDs. Looking forward, we anticipate using more high SNR spectra to perform detailed analysis of the H$\beta$/[O~\textsc{III}] region in LRDs, combining NIRCam and MIRI photometry to study the optical continuum properties of LRDs at higher redshifts, and conducting long-term spectroscopic monitoring to constrain the physical scales of the emission-line and continuum regions. We hope these efforts will lead to a clear understanding of the physical nature of the LRD population and the detailed properties of their gaseous structures in the near future.

\acknowledgments
We thank Zhiwei Pan for his useful discussions and suggestions. The authors would like to express their gratitude to the anonymous reviewer for his/her invaluable feedback and insightful suggestions.
This work is supported by the China Manned Space Program with grant no. CMS-CSST-2025-A06, the National Key R\&D Program of China No.2025YFF0510603, the National Natural Science Foundation of China (grant 12373009), the CAS Project for Young Scientists in Basic Research Grant No. YSBR-062, and the Fundamental Research Funds for the Central Universities. XW acknowledges the support by the Xiaomi Young Talents Program, and the work carried out, in part, at the Swinburne University of Technology, sponsored by the ACAMAR visiting fellowship. 
X.W. acknowledges the National Key R\&D Program of China (No. 2025YFA1614100).
These observations are associated with programs 1180, 1181, 1210, 1213, 1286, 1287, and 4233.
This work is based on observations made with the NASA/ESA/CSA
James Webb Space Telescope. The data were obtained from the
Mikulski Archive for Space Telescopes at the Space Telescope Science Institute, which is operated by the Association of Universities for Research in Astronomy, Inc., under NASA contract NAS 5-03127 for JWST.

\bibliography{JWST_LRD}{}
\bibliographystyle{aasjournal}
\end{document}